# A high dimensional delay selection for the reconstruction of proper Phase Space with Cross auto-correlation

**Sanjay Kumar Palit** [1*], **Sayan Mukherjee** [2] **and D K Bhattacharya** [3]

[1] *Mathematics Department, Calcutta Institute of Engineering and Management,*

*24/1A Chandi Ghosh Road, Kolkata-700040, INDIA.*

[2] *Mathematics Department, Shivanath Shastri College, 23/49 Gariahat Road,*

*Kolkata-700029, INDIA.*

[3] *Rabindra Bharati University, Kolkata-700050, INDIA.*

**Abstract.** For the purpose of phase space reconstruction from nonlinear time series, delay selection is one of the most vital criteria. This is normally done by using a general measure viz., mutual information (MI). However, in that case, the delay selection is limited to the estimation of a single delay using MI between two variables only. The corresponding reconstructed phase space is also not satisfactory. To overcome the situation, a high-dimensional estimator of the MI is used; it selects more than one delay between more than two variables. The quality of the reconstructed phase space is tested by shape distortion parameter (SD), it is found that even this multidimensional MI sometimes fails to produce a less distorted phase space. In this paper, an alternative nonlinear measure – cross auto-correlation (CAC) is introduced. A comparative study is made between the reconstructed phase spaces of a known three dimensional Neuro-dynamical model, Lorenz dynamical model and a three dimensional food-web model under MI for two and higher dimensions and also under cross auto-correlation separately. It is found that the least distorted phase space is obtained only under the notion of cross auto-correlation.

______________

[*] Corresponding author. Tel.: +919831004544.
E-mail address: sanjaypalit@ yahoo.co.in

# 1. Introduction

The entire focus of understanding nonlinear phenomena was concentrated, at the initial stage, on the experimentally observed irregular behaviour of systems that were predominantly obtained solely with computer simulations of appropriate mathematical models, e.g., mathematical model of Lorenz dynamical system and its unique three dimensional attractor [1]. Since its inception Takens' Embedding Theorem [2] has been used in time series analysis in many different fields ranging from system characterization and approximation of invariant quantities to prediction and noise-filtering. The Embedding Theorem asserts that although the true dynamics of a system may not be known, equivalent dynamics can be obtained under suitable conditions using time delays of a single time series, treated as a one-dimensional projection of the system trajectory. Thus most applications of the Embedding theorem deal with univariate time series, although measurements of more than one quantity related to the same dynamical system are often available. One of the first uses of multivariate embedding was in the context of spatially extended systems where embedding vectors were constructed from data representing the same quantity measured simultaneously at different locations [3, 4]. In nonlinear multivariate prediction, the prediction with local models on a space reconstructed from a different time series of the same system was studied in [5]. This study was extended in [6] by having the reconstruction utilizing all the observed time series. Multivariate embedding with the use of independent components analysis was considered in [7] and after that multivariate embedding with varying delay times was studied in [8, 9]. In fact, the dimension and the delay play the most important role for the phase space reconstruction, or embedding [10] of a time series. The state vectors, also called attractors, contain all the information available about the state of the process at a given time. They are also considered as the vectors that provide the most useful information in order to forecast the next state of the process. Since the publication of theoretical results about the dimension needed to reconstruct the structure of a series through an embedding in its phase space [2,10], many methods have been developed for estimating this dimension, such as correlation dimension [11], false neighbours [12], Box–Counting [10], minimum dimension [13], and such others. Concerning the selection of the delay, the common approaches are the use of the autocorrelation of the series, or the mutual information (MI) [14]. The goal is to select variables as phase space components that are as uncorrelated or independent as possible. Consequently each variable in the phase spaces provides useful information not provided by the other variables. However, the autocorrelation is a linear measure of dependency. It is only applied, when a linear model is built, but it does not provide the required information in a nonlinear context. The MI is a non-parametric measure of dependency between variables; it is applicable in both linear and nonlinear cases. Though MI is a criterion commonly used for delay selection, it suffers from two limitations. The first one is that MI is usually computed between two variables only and so it can detect the relations between two attractor values properly, but it fails to provide the information about high-dimensional relationships between all attractor values. The second limitation arises from the fact that only a unique delay is selected. The attractors are constructed using variables that are equally distributed in time (as multiples of the chosen delay). This approach artificially constrains the attractors in the sense that the delay is selected according to two-dimensional information, while the attractors themselves may be multi-dimensional. A more general approach would be to select different delays for the various variables taken into account in the attractors. The selection of these delays should ideally be performed in a space which has the same dimension as that of the attractors. A first attempt to solve this more general problem is to use a high-dimensional MI estimator, as the one provided in [15]. In [16], the high-dimensional MI estimator was used for better phase space reconstruction with multiple components of the solution vector of Lorenz system and Rossler system. Since Rossler system is a particular case of Lorenz system, so in this case the study was practically concentrated to Lorenz system only. Anyway MI is not a system dependent nonlinear measure, and so it cannot vary with different choices of dynamical systems. Hence it may not work for phase space reconstruction of dynamical systems, in general. Thus the results, which are found to be true for Lorenz or Rossler dynamical system, may not hold good for other systems. In fact, Neuro-dynamical model [17] is a good counter example. This is why we develop a new idea of high dimensional cross auto-correlation (CAC), which is system dependent. We expect that it would work for any dynamical system, in general to give better forms of phase spaces. But what are the criteria by which we can judge that our methods would give better reconstructed phase spaces? To answer this question, we make use of the shape distortion parameter (SD) [18]. More is the value of SD, less is the amount of shape distortion and hence better is the preservation of the dynamics of the attractor. So we find SD values under two and multi-dimensional MI and also under our new notion of cross auto-correlation and show that SD value is maximum under cross auto-correlation only. This proves that multi-dimensional CAC is best for phase space reconstruction.

## 2. Total Correlation and Cross Correlation

### 2.1. Total Correlation

In finding the equation of regression line, a 'line of best fit' is tried for the given two dimensional nonlinear points by the method of least squares. "Two dimensional nonlinear points" simply mean that they do not lie on a line or more precisely they cannot be approximated by a line of best fit. This ultimately determines linear regression coefficients. To find the corresponding nonlinear regression coefficients, a 'curve of best fit' is tried with the given two dimensional nonlinear points by the method of least squares. As they have a nonlinear trend, so the method of nonlinear correlation is to be applied to give proper judgment. The fact is that when two sets of data points have nonlinear (say parabolic) trend, ultimately by proper substitution, two sets of data points with linear trend are obtained and by the same correlation formula for linear trend, correlation coefficient is obtained even in this case of nonlinear trend.

The difficulty arises when we are to consider say, three time series of any suitable size. Naturally the question arises how to apply the standard correlation formula of two variables in the three dimensional case also. It is noted that if the points have a planar trend, then plane of regression may be tried. But this precisely means that variable of one of the time series must be connected by a linear function with the two other variables of the rest two time series. In this way a new data set of the approximated time series is obtained. Naturally we get two time series. One is the given one and the other one is the approximated one. So correlation formula for these two series is calculated and this is defined as the "total correlation" of the given three time series having planar trend.

When similar situation arises to find total correlation coefficients of *n* dimensional non-planer points, [that is when we want to compare *n* time series simultaneously] a 'hyper-plane of best fit' of the form $w = a_1 u_1 + a_2 u_2 + \ldots + a_{n-1} u_{n-1} + a_n$ is tried with the given points $(x_1(t), x_2(t), x_3(t), \ldots, x_n(t))$. Although *n* component points like $(x_1(t), x_2(t), x_3(t), \ldots, x_n(t))$ are always taken to compare *n* time series, but the number of such points is not *n*; this is why, while taking the summation for getting hyper-surface of best fit the summation is taken over any number suitable for computation.

With the help of these given *n* dimensional points and applying the method of least squares, values of $a_1, a_2, a_3, \ldots, a_n$ are determined, so that ultimately a plane of the form $w = f(u_1, u_2, u_3, \ldots, u_{n-1})$ is obtained, where *f* is a known linear function of $u_1, u_2, u_3, \ldots, u_{n-1}$. We call this new data set as $\{w'\}$.

The **total correlation coefficient** is defined as the linear correlation coefficient between $\{w'\}$ and the set $\{w''\}$, which is given by the $n^{th}$ component of the *n*-dimensional points like $(x_1(t), x_2(t), x_3(t), \ldots, x_n(t))$. The idea of **total correlation coefficient** is available in the literature [19].

## 2.2. Cross Correlation

A natural query is to know what happens if the three time series do not have a planar trend, but rather a curved surface trend. In this case, we cannot get two time series with linear trend by proper substitution as in the two dimensional case. So we are to extend the procedure for three time series with linear trend to three time series with nonlinear trend. Naturally a name other than "total correlation" is to be given. This new name is given as "cross-correlation" in general and "cross-auto-correlation" in particular. As a matter of fact, the points on the attractor are lying on a fractal set consisting of different non-planar, non-smooth points. So in such cases the notion of 'best plane fit' for *n* dimensional non-planar points is not workable. These points can be better approximated by a smooth hyper-surface, which we call a 'hyper-surface of best fit'. We first try to approximate the actual non-planar *n*-dimensional data points like $(x_1(t), x_2(t), x_3(t), \ldots, x_n(t))$ by fitting a smooth hyper-paraboloid of the form $w = a_1 u_1^{n-1} + a_2 u_2^{n-1} + a_3 u_3^{n-1} + \ldots + a_{n-1} u_{n-1}^{n-1} + a_n$. We then apply the method of least square and finally get the 'hyper-surface of best fit' given by $w = a_1 u_1^{n-1} + a_2 u_2^{n-1} + a_3 u_3^{n-1} + \ldots + a_{n-1} u_{n-1}^{n-1} + a_n$, where $a_1, a_2, a_3, \ldots, a_n$ are obtained from the following equations:

$$\left. \begin{aligned}
\Sigma w &= a_1 \Sigma u_1^{n-1} + a_2 \Sigma u_2^{n-1} + a_3 \Sigma u_3^{n-1} + \ldots + a_{n-1} \Sigma u_{n-1}^{n-1} + na_n \\
\Sigma u_1^{n-1} w &= a_1 \Sigma u_1^{2n-2} + a_2 \Sigma u_1^{n-1} u_2^{n-1} + a_3 \Sigma u_1^{n-1} u_3^{n-1} + \\
&\quad \ldots + a_{n-1} \Sigma u_1^{n-1} u_{n-1}^{n-1} + a_n \Sigma u_1^{n-1} \\
\Sigma u_2^{n-1} w &= a_1 \Sigma u_2^{n-1} u_1^{n-1} + a_2 \Sigma u_2^{2n-2} + a_3 \Sigma u_2^{n-1} u_3^{n-1} + \\
&\quad \ldots + a_{n-1} \Sigma u_2^{n-1} u_{n-1}^{n-1} + a_n \Sigma u_2^{n-1} \\
&\vdots \\
\Sigma u_{n-1}^{n-1} w &= a_1 \Sigma u_{n-1}^{n-1} u_1^{n-1} + a_2 \Sigma u_{n-1}^{n-1} u_2^{n-1} + a_3 \Sigma u_{n-1}^{n-1} u_3^{n-1} + \\
&\quad \ldots + a_n \Sigma u_{n-1}^{n-1} + a_{n-1} \Sigma u_{n-1}^{2n-2}
\end{aligned} \right\} \ldots (1)$$

Thus a hyper-surface of the form $w = f(u_1, u_2, u_3, \ldots, u_{n-1})$ is obtained, where $f$ is a known nonlinear function of $u_1, u_2, u_3, \ldots, u_{n-1}$. We call this new data set as $\{w'\}$.

The **cross correlation coefficient** is defined as the linear correlation coefficient between $\{w'\}$ and the set $\{w''\}$, which is given by the $n^{th}$ component of the $n$-dimensional points like $(x_1(t), x_2(t), x_3(t), \ldots, x_n(t))$.

The reason for choosing a hyper-paraboloid $w = a_1 u_1^{n-1} + a_2 u_2^{n-1} + a_3 u_3^{n-1} + \ldots + a_{n-1} u_{n-1}^{n-1} + a_n$ for this purpose is to minimize the computation time. In fact, the equation of the hyper-paraboloid contains one linear term and $(n-1)$ nonlinear terms, whereas other surfaces like hyper-ellipsoid or hyper-hyperboloid are represented by functions involving $n$ nonlinear terms. Naturally with later types of functions the calculation becomes more complex and unmanageable and takes more computational time. Moreover, there is no loss of generality in the choice of such simpler form of hyper-surface of best fit like a hyper-paraboloid, because for considering nonlinearity of the time series, we need to consider only small segments of the time series. So it makes no difference whether we consider equation of a hyper-paraboloid or a hyper-ellipsoid or a hyper-hyperboloid.

## 2.3. Cross Auto-correlation

It is a particular case of **cross correlation.** As we have compared $n$ time series simultaneously, so we have developed "cross-correlation" and "cross-auto-correlation" in $n$ dimensional spaces.

Let $x_1(t), x_2(t), x_3(t), \ldots, x_n(t)$ be the solution component of some $n$ dimensional dynamical system. To define **cross auto-correlation** for a single solution component $\{x(t)\}_{t=1}^{N}$ of the dynamical system with copies under same time-delay repeated and different time-delays, we proceed as follows:

### 2.3.1. Single time series with copies under same time-delay repeated

In this case, we first find the $n$ independent coordinates $(x(t), x(t+m), x(t+2m), \ldots, x(t+\overline{n-1}\,m))$ with time-delay $m$ as follows:

Firstly, the time series $\{x(t)\}_{t=1}^{N}$ is partitioned into $n$ groups, viz.,

$\{x(t)\}_{t=1}^{N-(n-1)m}, \{x(t)\}_{t=1+m}^{N-(n-2)m}, \{x(t)\}_{t=1+2m}^{N-(n-3)m}, \ldots, \{x(t)\}_{t=1+(n-1)m}^{N}$.

Thus we have $n$ sets of data in $n$ dimensional spaces given by $(u_{t,1}, u_{t,2}, u_{t,3}, \ldots, u_{t,n})$, where

$u_{t,1} = x(t), u_{t,2} = x(t+m), u_{t,3} = x(t+2m), \ldots, u_{t,n} = x(t+\overline{n-1}\,m)$   $t = 1, 2, 3, \ldots, (N-\overline{n-1}\,m)$.

Now $u_{t,1}, u_{t,2}, u_{t,3}, \ldots, u_{t,n-1}$ are substituted for $u_1, u_2, u_3, \ldots, u_{n-1}$ in the equation

$w = a_1 u_1^{n-1} + a_2 u_2^{n-1} + a_3 u_3^{n-1} + \ldots + a_{n-1} u_{n-1}^{n-1} + a_n$ of the best fitted surface to obtain a new time series $\{w_t\}_{t=1}^{N-(n-1)m}$. The **cross auto-correlation** between these $n$ time series $u_{t,1} = x(t), u_{t,2} = x(t+m), u_{t,3} = x(t+2m), \ldots, u_{t,n} = x(t+\overline{n-1}\,m)$ denoted by $r_{n,x}(m)$, is defined as the correlation between $\{u_{t,n}\}_{t=1}^{N-(n-1)m}$ and $\{w_t\}_{t=1}^{N-(n-1)m}$ with respect to the time-delay $m$ and is given by

$$r_{n,x}(m) = \frac{\sum_{t=1}^{N-(n-1)m} \{u_{t,n} - \overline{u_{t,n}}\} \cdot \{w_t - \overline{w_t}\}}{\sum_{t=1}^{N-(n-1)m} \{u_{t,n} - \overline{u_{t,n}}\}^2 \cdot \sum_{t=1}^{N-(n-1)m} \{w_t - \overline{w_t}\}^2}, \ldots (2)$$

where $m = 1, 2, 3, ....., (N-1)$ and $\overline{u_{t,n}}, \overline{w_t}$ are the mean of $u_{t,n}$ and $w_t$ respectively.

## 2.3.2. Single time series with copies under different time-delays

To find the independent coordinates

$$(x(t), x(t+m_1), x(t+m_1+m_2), ........., x(t+m_1+m_2+m_3+......+m_{n-1}))$$ for attractor reconstruction

with different time-delays $m_1, m_2, m_3, ......, m_{n-1}$, the time series is sub-divided into $n$ groups,

$$\{x(t)\}_{t=1}^{N-(m_1+m_2+m_3+....+m_{n-1})}, \{x(t)\}_{t=1+m_1}^{N-(m_2+m_3+....+m_{n-1})}, \{x(t)\}_{t=1+m_1+m_2}^{N-(m_3+....+m_{n-1})}, ....................., \{x(t)\}_{t=1+m_1+m_2+......+m_n}^{N}$$

Proceeding similarly as above, we define the **cross auto-correlation** between the time series

$$u_{t,1} = x(t), u_{t,2} = x(t+m_1), u_{t,3} = x(t+m_1+m_2), ........., u_{t,n} = x(t+m_1+m_2+........+m_{n-1})$$
$$t = 1, 2, 3, ....., (N-m_1+m_2+........+m_{n-1})$$

with respect to the time-delays $m_1, m_2, m_3, ......, m_{n-1}$, denoted by $r_{n,x}(m_1, m_2, m_3, ......, m_{n-1})$ as the correlation

between $\{u_{t,n}\}_{t=1}^{N-(m_1+m_2+....+m_{n-1})}, \{w_t\}_{t=1}^{N-(m_1+m_2+....+m_{n-1})}$ given by

$$r_{n,x}(m_1, m_2, ....., m_{n-1}) = \frac{\sum_{t=1}^{N-\sum_{l=1}^{n-1} m_l} \{u_{t,n} - \overline{u_{t,n}}\} \cdot \{w_t - \overline{w_t}\}}{\sqrt{\sum_{t=1}^{N-\sum_{l=1}^{n-1} m_l} \{u_{t,n} - \overline{u_{t,n}}\}^2 \cdot \sum_{t=1}^{N-\sum_{l=1}^{n-1} m_l} \{w_t - \overline{w_t}\}^2}}, \quad .... (3)$$

where $m_1, m_2, ......, m_{n-1} = 1, 2, 3, ....., (N-1)$.

Here we must note the following points:

(i) Cross auto-correlation (CAC) is a nonlinear measure.

(ii) As $w = a_1 u_1^{n-1} + a_2 u_2^{n-1} + a_3 u_3^{n-1} + ..... + a_{n-1} u_{n-1}^{n-1} + a_n$ is a family of hyper-paraboloid, so different members of the family are determined by different sets of values of $a_1, a_2, a_3, ........., a_n$. Again different sets of values of $a_1, a_2, a_3, ........., a_n$ are found from the equations given by (1) for different set of points $(x_1(t), x_2(t), x_3(t), ....., x_n(t))$, which are actual points on the attractor itself. As the dynamical systems vary, their attractors satisfy different $(x_1(t), x_2(t), x_3(t), ....., x_n(t))$ and so values of $a_1, a_2, a_3, ........., a_n$ are also found to be different.

Thus our nonlinear measure also varies from dynamical system to dynamical system; it is actually system dependent.

### 2.3.3. Multiple time series with copies under same time-delay repeated

Let $(x_1(t), x_2(t), x_3(t), ....., x_n(t))$ be the solution component of some dynamical system.

In this case, the independent coordinates are taken as $x(t), x(t+m), x(t+2m), ........., x(t+\overline{n-1}\,m)$ $(t = 1, 2, 3, ...., N-2m)$, under single time-delay $m$, such that

$$x(t) = x_{i_1}(t),\ x(t+m) = x_{i_2}(t+m),\ x(t+2m) = x_{i_3}(t+2m), ........., x(t+\overline{n-1}\,m) = x_{i_n}(t+\overline{n-1}\,m)$$

where $i_1, i_2, i_3, ........., i_n \in \{1, 2, 3, ....., n\}$. The cross auto-correlation between these $n$ time series is then obtained by (2).

### 2.3.4. Multiple time series with copies under different time-delays

For different time-delays $m_1, m_2, m_3, ........., m_{n-1}$ the independent coordinates are taken as $x(t), x(t+m_1), x(t+m_1+m_2), ......., x(t+m_1+m_2+......+m_{n-1})$,

$t = 1, 2, 3, ......, (N - \sum_{l=1}^{n-1} m_l)$, such that

$$x(t) = x_{i_1}(t),\ x(t+m_1) = x_{i_2}(t+m_1),\ x(t+m_1+m_2) = x_{i_3}(t+m_1+m_2), ....., x(t+\sum_{l=1}^{n-1} m_l) = x_{i_n}(t+\sum_{l=1}^{n-1} m_l),$$

where $i_1, i_2, i_3, ........., i_n \in \{1, 2, 3, ....., n\}$. Finally, the cross auto-correlation between these $n$ time series is obtained by (3).

# 4. Determination of suitable time-delay /delays for attractor reconstruction with cross auto- correlation

## 4.1. Under same time-delay repeated

To obtain the suitable time-delay $m$, the cross auto-correlation $r_{n,x}(m)$ under same time-delay repeated given by equation (2) for different values of $m$ is plotted against $m$. This is known as two dimensional correlogram diagram.

To get the independent coordinates, $(x(t), x(t+m), x(t+2m), \ldots, x(t+\overline{n-1}\,m))$, $t = 1, 2, 3, \ldots, (N - \overline{n-1}\,m)$ we choose that value of $m$ for which $r_{n,x}(m)$ comes nearest to zero for the first time in the two dimensional correlogram diagram. The attractor in $n$ dimension is then reconstructed with these independent coordinates of the time series $\{x(t)\}_{t=1}^{N}$.

## 4.2. Under different time-delays

To obtain the suitable time-delays $m_1, m_2, m_3, \ldots, m_{n-1}$, the cross auto-correlation $r_{n,x}(m_1, m_2, m_3, \ldots, m_{n-1})$ given by equation (3) for different values of $m_1, m_2, m_3, \ldots, m_{n-1}$ is plotted against $m_1, m_2, m_3, \ldots, m_{n-1}$. This is known as three dimensional correlogram diagram. To get the independent coordinates $(x(t), x(t+m_1), x(t+m_1+m_2), \ldots, x(t+\sum_{l=1}^{n-1} m_l))$, $t = 1, 2, 3, \ldots, (N - \sum_{l=1}^{n-1} m_l)$, we choose that value of $(m_1, m_2, m_3, \ldots, m_{n-1})$ for which $r_{n,x}(m_1, m_2, m_3, \ldots, m_{n-1})$ comes nearest to zero for the first time in the three dimensional correlogram diagram. The attractor in $n$ dimensional space is then reconstructed with these independent coordinates of the time series $\{x(t)\}_{t=1}^{N}$.

## 5. A known Neuro-dynamical model and its chaotic phase space

Consider a known Neuro-dynamical model [17] given by the following differential equations:

$$\frac{dx_1}{dt} = \left[1+\exp\{-\beta_1(w_{21}x_2 + w_{31}x_3 - \theta_1)\}\right]^{-1} - \alpha_1 x_1$$
$$\frac{dx_2}{dt} = \left[1+\exp\{-\beta_2(x_1 - \theta_2)\}\right]^{-1} - \alpha_2 x_2 \qquad (4)$$
$$\frac{dx_3}{dt} = \left[1+\exp\{-\beta_3(x_1 - \theta_3)\}\right]^{-1} - \alpha_3 x_3$$

with the initial condition $x_1(1) = 0.8, x_2(1) = 0.5, x_3(1) = 0.1$ and the parameters are given by

$w_{21} = 1$, $w_{31} = -6.2$, $\alpha_1 = 0.62$, $\alpha_2 = 0.42$, $\alpha_3 = 0.1$, $\beta_1 = 7$, $\beta_2 = 7$, $\beta_3 = 13$, $\theta_1 = 0.5$, $\theta_2 = 0.3$, $\theta_3 = 0.7$.

Under these values of the parameters, the system gives a chaotic phase space shown by fig.1.

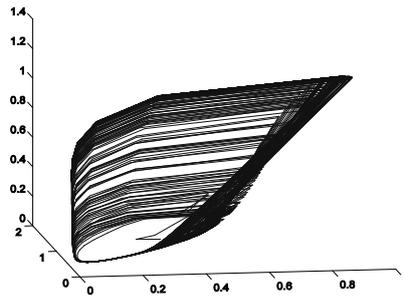

**Fig.1: Chaotic Phase space for the Neuro-dynamical model.**

Solving the above Neuro-dynamical model for $x_1, x_2, x_3$, we get three time series $\{x_1(t)\}_{t=1}^{10000}, \{x_2(t)\}_{t=1}^{10000}, \{x_3(t)\}_{t=1}^{10000}$ as solutions. We next calculate the Lyapunov exponent for each of $\{x_1(t)\}_{t=1}^{10000}, \{x_2(t)\}_{t=1}^{10000}, \{x_3(t)\}_{t=1}^{10000}$ and observe that the average Lyapunov exponent is positive. This indicates that phase space reconstruction is possible for the aforesaid Neuro-dynamical model from any one of $x_1, x_2, x_3$.

## 6. Suitable delay selection for the phase space reconstruction

### 6.1. Phase space reconstruction from single component of the solution vector under AMI

We first choose the solution component $\{x_3(t)\}_{t=1}^{10000}$ arbitrarily for the reconstruction of three dimensional phase space. Fig.2 shows the plot of (AMI) versus time-delay $m$ for $\{x_3(t)\}_{t=1}^{10000}$.

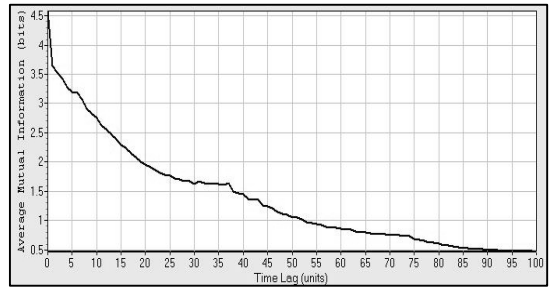

**Fig.2: Average Mutual Information (AMI) Vs. Time-delay.**

It is seen from fig.2 that AMI comes nearer to zero for the first time, when $\tau = 30$. The corresponding three dimensional phase space with $\tau = 30$ repeated is given by fig.3.

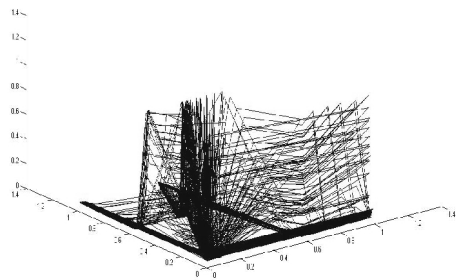

**Fig.3: Three dimensional Phase space of the Neuro-dynamical model with $\tau = 30$ repeated.**

Obviously this differs much from the original phase space given by fig.1. We also apply the present method for the solution components $\{x_1(t)\}_{t=1}^{10000}$, $\{x_2(t)\}_{t=1}^{10000}$ of the Neuro-dynamical model, but the phase space reconstruction does not improve. In fact, in the later cases the reconstructed phase spaces are even worse than the present one.

## 6.2. Phase space reconstruction from multiple components of the solution vector under multi-dimensional MI with same time-delay repeated

We now use multi-dimensional Mutual Information [18] to find suitable time-delay to reconstruct the phase space of the Neuro-dynamical model from multiple components of its solution vector. The phase spaces are reconstructed from all possible combinations of the solution component with same time-delay repeated. All the phase spaces, when judged visually, show that their best form is given as in fig.4.

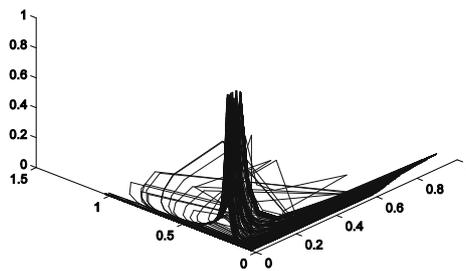

**Fig.4: The best form of reconstructed phase space with different combinations of solution components with same time-delay repeated.**

Fig.4 shows a bit improvement over fig.3 in the sense that in the present case we get a better dense region with lesser number of outliers but the improvement is not at all remarkable.

## 6.3. Phase space reconstruction from multiple components of the solution vector under multi-dimensional MI with different time-delays

We next use multi-dimensional Mutual Information [18] to find proper time-delays to reconstruct the phase space of the Neuro-dynamical model from multiple components of its solution vector with different time-delays. All the phase spaces reconstructed under multidimensional Mutual Information, when judged visually, show that their best form is given as in fig.5.

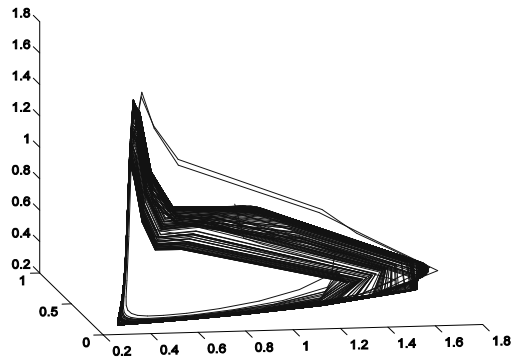

**Fig.5: The best form of reconstructed phase space with different combinations of solution components with different time-delays.**

Fig.5 shows that we are getting a comparatively better dense region with lesser number of outliers as compared to the previous ones. But it still differs much from the actual attractor given by fig.1. More improvements remain to be achieved. But this cannot be done by existing methods as mentioned above.

**6.4. Phase space reconstruction under cross auto-correlation from single component of the solution vector**

We consider each of the solution components separately and try to reconstruct the phase space. For the sake of illustration, we present here only the reconstructed phase space from the solution component $\{x_2(t)\}_{t=1}^{10000}$. The quality of the two other phase spaces reconstructed from the other two solution components of the Neuro-dynamical model is found to be even worse than this.

### 6.4.1. Under same time-delay repeated

Since the present dynamical system is three dimensional, Cross auto-correlations $r_{n,x_2}(m)$ given by equation (2) with $n = 3$ is used to find the independent coordinates $(x_2(t), x_2(t+m), x_2(t+2m))$, $t = 1,2,3,...,N-2m$, where $m$ is found to be 7 (Fig.6).

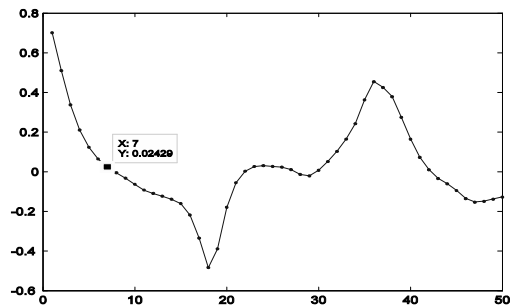

**Fig.6: Two dimensional Correlogram for the solution component** $\{x_2(t)\}_{t=1}^{10000}$.

The attractor reconstructed with $m = 7$ is shown by fig.7.

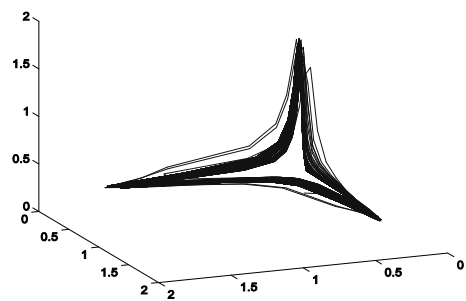

**Fig.7. Three dimensional reconstructed phase space under cross auto-correlation with $m = 7$ repeated.**

### 6.4.2. Under different time-delays

We now reconstruct the phase space of the Neuro-dynamical model by considering the same solution component $\{x_2(t)\}_{t=1}^{10000}$ of its solution vector under different time-delays $m_1$ and $m_2$. For this purpose, the cross auto-correlation given by (3) with $n = 3$ is used to evaluate the values of $r_{3,x_2}(m_1, m_2)$ for different values of $m_1$ and $m_2$, which are then plotted against $m_1$ and $m_2$ to obtain the three dimensional correlogram shown in fig.8.

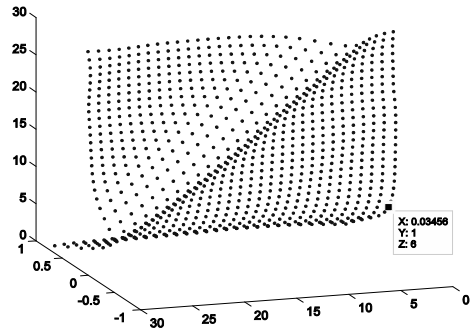

**Fig.8: Three dimensional Correlogram for $\{x_2(t)\}_{t=1}^{10000}$ under different time-delays.**

It is observed from fig.8 that the value of the cross auto-correlation $r_{3,x_2}(m_1, m_2)$ comes nearer to zero for the first time when $m_1 = 1$ and $m_2 = 6$. The reconstructed phase space is shown in fig.9.

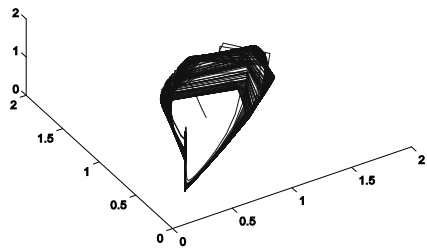

**Fig.9: Three dimensional reconstructed phase space with $m_1 = 1, m_2 = 6$.**

It is observed from fig.7 and fig.9 respectively that the phase spaces reconstructed by our newly proposed nonlinear method with same time-delay repeated and different time-delays exhibit orbits, which are almost dense. So they may be considered as very good attractors. In fact, there is a trend of improvement from fig.4 to fig.5 and then from fig.7 to fig.9. Thus so far as reconstruction is concerned from single component, our measure is even better than multi-dimensional MI with same time-delay repeated and different time-delays. But none of the figures (Fig.7 and Fig.9) resembles the actual phase space given by fig.1. So cross auto-correlation is also not much effective if phase space reconstruction is tried from a single component.

**6.5. Phase space reconstruction under cross auto-correlation from multiple components of the solution vector**

**6.5.1. Under same time-delay repeated**

To reconstruct the phase space of the above Neuro-dynamical model using multiple solution components, we first obtain the independent coordinates $(x(t), x(t+m), x(t+2m))$ where $m$ is the single time-delay repeated and

$$x(t) = x_{i_1}(t), x(t+m) = x_{i_2}(t+m), x(t+2m) = x_{i_3}(t+2m), i_1, i_2, i_3 \in \{1, 2, 3\}.$$

The best form of the phase space reconstructed under this method is shown in fig.10.

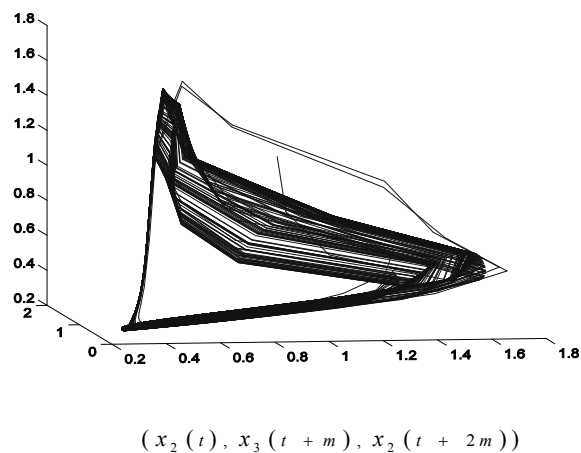

$(x_2(t), x_3(t+m), x_2(t+2m))$

**Fig.10: Reconstructed phase space of the aforesaid Neuro-dynamical model with different under same time-delay ($m$=2) repeated.**

In fact, it is evident from fig.10 that the reconstructed phase space with the combinations $x_2$, $x_3$, $x_2$ may be considered as a very good attractor for the Neuro-dynamical model with single time-delay, as it exhibits complete dense orbits. Also this reconstructed phase space has some similarity with the actual phase space of the Neuro-dynamical model given by fig.1.

### 6.5.2. Under different time-delays

We now reconstruct phase spaces for the Neuro-dynamical model by choosing the independent coordinates $(x(t), x(t+m_1), x(t+m_1+m_2))$, for two different time-delays $m_1$, $m_2$, where

$$x(t) = x_{i_1}(t), x(t+m_1) = x_{i_2}(t+m_1), x(t+m_1+m_2) = x_{i_3}(t+m_1+m_2), i_1, i_2, i_3 \in \{1, 2, 3\}.$$

The best form of the phase space reconstructed under this method is shown in fig.11.

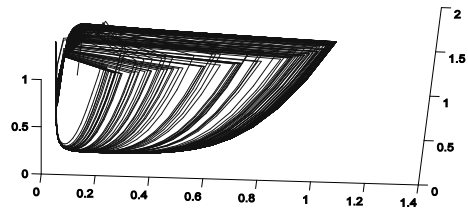

**Fig.11: The best form of reconstructed phase spaces of the Neuro-dynamical model with different combinations of the components of its solution vector under different time-delays ($m_1$=1, $m_2$=3).**

In fact, the reconstructed phase space with the combination $x_3$, $x_2$, $x_1$ under different time-delays [fig.11] has remarkably improved compared to that obtained in case of same time-delay repeated [fig.10]. Fig.11 contains lesser number of outliers and also a better dense orbit as compared to those obtained under same time-delay repeated [fig.10]. Also it is very much similar to the actual phase space [fig.1].

# 7. Results and Discussions

To establish the supremacy of our newly proposed notion of CAC, we first compute the value of the shape distortion parameter (SD) in each case under the existing and our newly proposed notion for the aforesaid Neuro-dynamical model.

| Method Applied to find suitable time-delay/time-delays | Average Mutual Information (AMI) | High dimensional MI with same time-delay repeated | High dimensional MI with different time-delays | High dimensional CAC with same time-delay repeated | High dimensional CAC with different time-delays |
|---|---|---|---|---|---|
| Independent coordinates for best phase space reconstruction | $(x_3(t), x_3(t+m), x_3(t+2m))$ | $(x_1(t), x_3(t+m), x_1(t+2m))$ | $(x_2(t), x_1(t+m_1), x_2(t+m_2))$ | $(x_2(t), x_3(t+m), x_2(t+2m))$ | $(x_3(t), x_2(t+m_1), x_1(t+m_1+m_2))$ |
| Shape Distortion Parameter (SD) | 0.0014 | 0.0125 | 0.165003 | 0.367 | 0.4242 |

Table.1. Tabulation of Shape distortion parameter values for Neuro-dynamical model.

The SD values obtained for the best form of reconstructed phase spaces under each of the existing methods and also under the newly proposed method are tabulated in table.1.

From the above table, following decisions are made:

(i) The results obtained by visual inspection have been confirmed by the shape distortion parameter SD. Table.1 maintains a steadily increasing trend in the values of SD.

(ii) The highest SD value occurs in the 6[th] (last) column only. This proves that the corresponding phase space given by fig. 11 is least distorted in shape. Hence the dynamics is best preserved in this case.

Thus multi-dimensional cross auto-correlation with multiple solution components is the best tool for reconstruction of phase space.

Now to examine the applicability of this newly proposed nonlinear measure, we carry out phase space reconstruction of the famous Lorenz's model [21] and a known food-web model [22] under both existing methods and the newly proposed nonlinear measure – CAC.

The SD values obtained for the best form of reconstructed phase spaces of Lorenz's model and the food-web model under each of the existing methods and also under the newly proposed method are tabulated in table.2 and table.3 respectively.

| Method Applied to find suitable time-delay/time-delays | Average Mutual Information (AMI) | High dimensional MI with same time-delay repeated | High dimensional MI with different time-delays | High dimensional CAC with same time-delay repeated | High dimensional CAC with different time-delays |
|---|---|---|---|---|---|
| Independent coordinates for best phase space reconstruction | $(x_1(t), x_1(t+m), x_1(t+2m))$ | $(x_3(t), x_1(t+m), x_2(t+2m))$ | $(x_3(t), x_1(t+m_1), x_1(t+m_2))$ | $(x_3(t), x_1(t+m), x_2(t+2m))$ | $(x_3(t), x_1(t+m_1), x_1(t+m_1+m_2))$ |
| Shape Distortion Parameter (SD) | 0.5838 | 0.6298 | 0.8896 | 0.7625 | 0.8994 |

Table.2. Tabulation of Shape distortion parameter values for Lorenz's model.

| Method Applied to find suitable time-delay/time-delays | Average Mutual Information (AMI) | High dimensional MI with same time-delay repeated | High dimensional MI with different time-delays | High dimensional CAC with same time-delay repeated | High dimensional CAC with different time-delays |
|---|---|---|---|---|---|
| Independent coordinates for best phase space reconstruction | $(x_1(t), x_1(t+m), x_1(t+2m))$ | $(x_1(t), x_2(t+m), x_3(t+2m))$ | $(x_3(t), x_1(t+m_1), x_1(t+m_2))$ | $(x_1(t), x_1(t+m), x_3(t+2m))$ | $(x_1(t), x_3(t+m_1), x_1(t+m_1+m_2))$ |
| Shape Distortion Parameter (SD) | 0.1015 | 0.2380 | 0.6015 | 0.6520 | 0.7805 |

Table.3. Tabulation of Shape distortion parameter values for the food-web model.

It is found from both of table.2 and table.3 that the values of the shape distortion parameter (SD) maintains almost same trend as Neuro-dynamical model. In fact, the SD value is greatest in column 6 of table.2 and table.3, obtained under our newly proposed nonlinear measure – CAC by considering multiple solution components and different time-delays. This not only confirms the applicability of the nonlinear measure – CAC for the best possible phase space reconstruction of any dynamical system, but also indicates the necessity of considering multiple solution components [18] and the necessity of using different time- delays [20] for a better phase space reconstruction.

## 8. Conclusion

In this article, advancement to the existing approaches of single/different time-delay(s) selection for phase space reconstruction from single/multiple time series has been proposed. It was found that phase space reconstruction for the Neuro-dynamical model, which was impossible by AMI, almost impossible by high dimensional Mutual Information (MI) method with same time-delay repeated, quite possible by high dimensional Mutual Information (MI) method with different time-delays, become possible under our newly proposed nonlinear measure CAC especially with different time-delays. The value of the shape distortion parameter given by table.1 establishes that the phase space reconstructed under the notion of CAC with different time-delays have least shape distortion from the original phase space. The same trend was observed in case of the famous Lorenz model and a known food-web model. In fact, our newly proposed notion of CAC is a system dependent measure and so it is expected to work in phase space reconstruction for any dynamical systems. As we are making use of the shape distortion parameter (SD) that does not depend on the dimension of the phase space, naturally if we consider higher dimensional phase spaces of dimension greater than three, where visual inspection is no longer workable, then also greatest SD value would ensure the best phase space. Hence CAC may be used in higher dimensional phase space reconstruction. Thus, it may be concluded that CAC is one of the best nonlinear measure for phase space reconstruction, invented so far.

———————————